# VQkit: A Unified and Extensible Framework for VASP and Quantum ESPRESSO Pre- and Post-Processing

*Rajkamal Anand* [a, *], *Seunghwan Kwon* [b], *Gun Jin Yun* [a, b, c]

[a] *Institute of Advanced Aerospace Technology, Seoul National University, Gwanak-gu Gwanak-ro 1, Seoul 08826, Republic of Korea*

[b] *Department of Aerospace Engineering, Seoul National University, Gwanak-gu Gwanak-ro 1, Seoul, 08826, Republic of Korea*

[c] *Institute of Engineering Research, Seoul National University, Gwanak-gu Gwanak-ro 1, Seoul, 08826, Republic of Korea*

[*]Corresponding author: rajkamalloyola@gmail.com (Rajkamal A)

## Abstract

VQkit is a Python-based computational toolkit developed to streamline the pre- and post-processing of density functional theory calculations performed using Vienna Ab initio Simulation Package (VASP) and Quantum ESPRESSO (QE). The framework employs a unified representation of lattice vectors, atomic coordinates, chemical species, and positional constraints, enabling cross-code structure conversion, geometry manipulation, crystallographic reconstruction, and code-specific input generation. Its pre-processing capabilities include symmetry-dependent reciprocal-space path generation, atomic constraints, and automated preparation of simulation inputs. The post-processing framework provides calculation convergence inspection, extraction of total energies and Fermi levels, electronic band-structure and density-of-states analysis, orbital-resolved electronic-structure processing, charge-density analysis, and evaluation of application-specific materials properties. The computational implementation is evaluated through elastic-property benchmarks for HfC, ZrC, $ZrO_2$, Ti, and diamond, showing close agreement between VQkit and VASPKIT for VASP-based calculations, with differences observed in the corresponding QE results. By integrating cross-code interoperability, automated simulation preparation, standardized post-processing, and quantitative property evaluation within a common computational environment, VQkit reduces workflow fragmentation and facilitates consistent pre- and post-processing of first-principles calculations.

**Program title:** VQkit: VASP and Quantum ESPRESSO Interface Kit,

**Licensing provisions:** MIT License,

**Programming language:** Python

**Nature of problem:** Differences in VASP and Quantum ESPRESSO input/output formats complicate unified preprocessing, post-processing, and materials-property analysis.

**Solution method:** VQkit provides a unified Python-based interface for structure preparation, input generation, output analysis, visualization, and materials-property evaluation using VASP and Quantum ESPRESSO.

## 1. Introduction

Density functional theory (DFT)[1,2] is a cornerstone of modern computational materials science, providing a quantum-mechanical framework for describing and predicting a wide range of material properties[3], including structural[4], electronic[5], magnetic[6], mechanical[7], optical[8], and catalytic[9,10] characteristics. Progress in electronic-structure methods, numerical techniques, and high-performance computing has considerably expanded the scope of DFT, enabling first-principles investigations of increasingly complex materials as well as high-throughput screening across large materials spaces. A broad range of electronic-structure packages is currently available under different licensing models. For example, *Vienna Ab initio Simulation Package (VASP)*[11], *WIEN2k*[12], and Cambridge Serial Total Energy Package (*CASTEP*)[13] are distributed under licensing agreements, whereas *Quantum ESPRESSO (QE)*[14], *ABINIT*[15], and the Spanish Initiative for Electronic Simulations with Thousands of Atoms (*SIESTA)*[16] are available as open-source software. For many widely used plane-wave DFT packages, including *VASP* and QE, practical calculations involve preparing code-specific input files, executing the electronic-structure calculation, and subsequently processing the generated outputs to extract structural, energetic, electronic, and other physically relevant quantities.

Several electronic-structure packages are supported by dedicated utilities or closely associated software environments that facilitate calculation setup, workflow management, and post-processing. *ABINIT* is complemented by native post-processing utilities and the *AbiPy*[17] framework, which provides capabilities for structure conversion, high-symmetry *k*-path generation, workflow automation, and electronic-structure analysis. *WIEN2k* includes the *w2web*[18] environment for structure preparation, input initialization, calculation execution, and analysis of properties such as the density of states (DOS). *SIESTA* is accompanied by utilities for electronic structure, and related analyses, while the *sisl[19]* framework provides additional capabilities for processing *SIESTA* calculations. *CASTEP* is similarly supported by graphical and auxiliary environments, including *BIOVIA* Materials Studio[20] and complementary electronic-structure analysis tools. These developments illustrate the important role of integrated or code-specific support environments in practical DFT workflows.

*VASPKIT* [21], which supports structure preparation, input generation, and post-processing for *VASP*, and PWgui, which provides a graphical interface for preparing *QE* input files, alongside dedicated *QE* utilities for electronic-structure analysis. More recently, the *QE* app within Automated Interactive Infrastructure and DAtabase (*AiiDA) lab* has introduced a graphical, workflow-oriented environment for automated *AiiDA*-based calculations and analyses. These tools substantially improve the usability of their respective electronic-structure environments; however, they remain primarily organized around code-specific workflows and conventions. Native *QE* also relies on multiple input files, executables, and auxiliary tools rather than a single interface spanning the complete preparation and post-processing workflow. Broader materials-design platforms such as *JARVIS*[22] also support multiple simulation packages, including *VASP* and *QE*. However, their scope is not centered on a lightweight framework for equivalent day-to-day preprocessing, structural manipulation, direct *VASP-QE* conversion, constraint handling, and post-processing across these two codes.

Recent advances in large language models (*LLMs*)[23] and Artificial Intelligence (*AI*)-assisted coding tools[24] have enabled rapid generation of scripts and support for computational pre- and post-processing. However, many advanced *AI* services require paid subscriptions or Application Programming Interface (*API*) access, while their outputs may depend on model versions, prompts, and conversational context. General-purpose *LLMs* may also generate scientifically inconsistent or incorrect simulation inputs, particularly for complex workflows. Domain-specific software therefore remains valuable for operations that should be deterministic, inspectable, and reproducible.

Hence, we developed VQkit, a file-based command-line framework for unified *VASP* and *QE* pre- and post-processing (**Figure 1**). Its distinguishing feature is a common internal representation in which the lattice, atomic configuration, coordinates, constraints, and calculation state are treated independently of code-specific syntax. This enables direct *VASP-QE* structure and input conversion, code-independent structural manipulation, layer- and atom-based constraints, supercell and coordinate transformations, and generation of equivalent inputs for both electronic-structure codes without manually rebuilding the model. Beyond conventional preprocessing, VQkit provides a unified framework for analyzing electronic band structures, DOS, PDOS, spin-resolved electronic states, and band gaps. It also integrates application-oriented modules for batteries, electrocatalysis, mechanical properties, phonon-derived descriptors, and supercapacitors, supporting both *VASP* and *QE* within the same computational environment. The following sections describe the common data model, implementation, numerical validation examples, present limitations and future development priorities.

## 2. Software architecture and common data representation

A central design principle of VQkit is the separation of general workflow operations from code-specific input and output handling. Structural information, atomic coordinates, lattice parameters, reciprocal-space definitions, calculation settings, and selected analysis quantities are organized through a common internal representation before being passed to the corresponding *VASP* or *QE* interface. This abstraction allows the same structural and geometrical operations to be applied independently of the underlying simulation engine and reduces duplication of equivalent functionality across different code-specific modules.

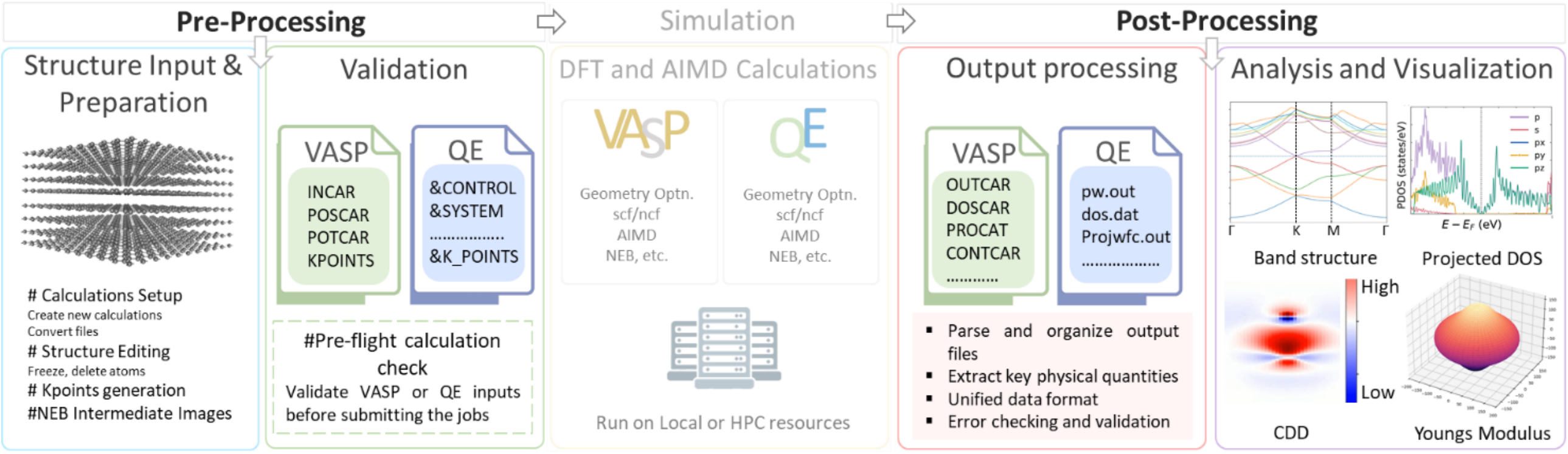

***Figure 1.*** *Workflow of VQkit for unified pre- and post-processing of VASP and QE calculations.*

Within this framework, structure-related operations such as coordinate conversion, atom selection, layer identification, height-based filtering, atom deletion, and positional constraints are performed at the common-data level. The processed information is then translated into the syntax required by the selected DFT code. Structural information, atomic coordinates, lattice parameters, reciprocal-space definitions, and supported calculation settings are organized through a common internal representation before being passed to the corresponding *VASP* or *QE* interface. Similarly, lattice and atomic information can be converted between *POSCAR*-like and *QE*-compatible representations while preserving the same physical structure.

The use of a common representation also simplifies post-processing. Quantities extracted from code-specific output files are mapped into standardized variables using consistent units, coordinate conventions, and energy references, where applicable, before subsequent analysis or visualization. *VASP* and *QE* report several physical quantities using different units and coordinate conventions. For example, total energies are typically reported in electronvolts ($eV$) by *VASP* and Rydberg ($Ry$) units by *QE*, whereas atomic positions may be expressed in Cartesian or fractional coordinates using angstroms ($\AA$), Bohr radii, or lattice-dependent units. Stress tensors are commonly reported in kilobars ($kbar$), while electronic eigenvalues are generally available in $eV$ in the relevant output files. To ensure consistent post-processing, VQkit converts these quantities into common representations, including total energies and electronic eigenvalues in $eV$, atomic positions in $\AA$, and stress tensors in $GPa$. The energy conversion follows 1 $Ry = 13.6\ eV$, while stress values are converted using 1 $kbar = 0.1\ GPa$. These conversions enable physical quantities extracted from *VASP* and *QE* outputs to be processed using consistent numerical units within the VQkit framework.

As a result, operations such as convergence inspection, energy extraction, band-energy processing, and DOS analysis can follow similar internal workflows even though the original output formats differ substantially between *VASP* and *QE*. This layered organization forms the basis of the cross-code functionality of VQkit and provides a practical foundation for incorporating additional simulation engines in future versions.

## 3. Implementation and computational procedures

### 3.1. Physical representation and software dependencies

VQkit is implemented in *Python*, with *NumPy*[25] used for numerical operations and *Matplotlib*[26] for data visualization. Structural parsing, coordinate transformations, structure manipulation, and *VASP-QE* conversion routines are implemented directly within VQkit. For crystallographic symmetry analysis and standardized high-symmetry band-path generation, VQkit interfaces with *spglib*[27] and *SeeK-path*[28], respectively. The following sections describe the physical representations and computational procedures used to provide a common treatment of crystal structures and reciprocal-space quantities across *VASP* and *QE*.

### 3.2. Pre-processing and its workflow generation

This module provides a unified workflow for preparing and validating calculations for *VASP* and *QE*. Rather than treating input preparation as a collection of code-specific file operations, the module organizes the workflow around the underlying structural model and computational task. Users can create new calculations, convert crystallographic information files (CIF) into *VASP* or *QE* inputs, and regenerate input structures from the final geometry of completed calculations, thereby facilitating sequential workflows such as relaxation followed by self-consistent-field or electronic-structure calculations. Geometry manipulation is integrated directly into the same framework: selected atoms can be constrained according to index, chemical species, layer, or spatial position, or removed from the structure when constructing surfaces, interfaces, defects, and adsorbate models. Before job submission, a pre-flight validation routine checks the consistency of the structural and calculation inputs to identify common setup errors. VQkit also manages code-specific pseudopotential resources through automatic *POTCAR* construction for *VASP* and configurable pseudopotential-library locations for both *VASP* and *QE*. For electronic band-structure calculations, the module determines the crystallographic cell symmetry and generates the corresponding high-symmetry reciprocal-space path in the format required by either code.

### 3.2.1. Unified structural representation

VQkit separates the physical description of a periodic structure from the syntax used by a particular electronic-structure code. Structural information obtained from *VASP*, *QE*, and *CIF* files are first mapped onto a common internal representation containing the lattice vectors, atomic species, atom ordering, and atomic coordinates. This representation enables subsequent structural operations to be performed on the underlying physical geometry rather than directly on code-specific text files.

A periodic simulation cell is defined by three real-space lattice vectors, *a*, *b*, and *c*. In VQkit, these vectors are represented as column vectors of the lattice matrix '*A*'. Writing their Cartesian components explicitly, the lattice geometry can be expressed as

$$A = [a \;\; b \;\; c] = \begin{pmatrix} a_x & b_x & c_x \\ a_y & b_y & c_y \\ a_z & b_z & c_z \end{pmatrix} \tag{1}$$

where the three columns of '*A*' correspond to the real-space lattice vectors. For example, consider a cubic unit cell with lattice constant $a_0 = 3$ Å. The lattice vectors are $a = (3, 0, 0)$, $b = (0, 3, 0)$, and $c = (0, 0, 3)$ Å, giving

$$A = \begin{pmatrix} 3 & 0 & 0 \\ 0 & 3 & 0 \\ 0 & 0 & 3 \end{pmatrix} \text{Å} \tag{2}$$

They define the periodic simulation cell, whereas the atomic coordinates specify the positions of atoms within the cell. For atom $i$, the fractional-coordinate vector is

$$s_{\mathrm{i}} = \begin{pmatrix} u_{\mathrm{i}} \\ v_{\mathrm{i}} \\ w_{\mathrm{i}} \end{pmatrix} \tag{3}$$

and the corresponding Cartesian-coordinate vector is obtained from

$$r_{\mathrm{i}} = As_{\mathrm{i}} = u_{\mathrm{i}}a + v_{\mathrm{i}}b + w_{\mathrm{i}}c \tag{4}$$

with the inverse transformation

$$s_{\mathrm{i}} = A^{-1}r_{\mathrm{i}} \tag{5}$$

For example, an atom at fractional coordinates (0.5, 0.5, 0.5) in the cubic cell of Eq. (2) is located at $r = 0.5(3, 0, 0) + 0.5(0, 3, 0) + 0.5(0, 0, 3) = (1.5, 1.5, 1.5)$ Å. Eqs. (4) and (5) therefore provide a common geometrical basis for normalizing structures expressed using different coordinate conventions, including *VASP* Direct or Cartesian coordinates and *QE* crystal- or Cartesian-based coordinates.

### 3.2.2. Crystallographic reconstruction and CIF-based input generation

For CIF-based input generation, VQkit reconstructs the periodic structure from the crystallographic information contained in the input file. A CIF typically provides the lattice parameters $a$, $b$, and $c$, the interaxial angles $\alpha$, $\beta$, and $\gamma$, chemical species, fractional atomic coordinates, and, when available, space-group symmetry information. The lattice parameters are related to the real-space basis vectors through

$$a = |\boldsymbol{a}|, \quad b = |\boldsymbol{b}|, \quad c = |\boldsymbol{c}| \tag{6a}$$

$$\alpha = \cos^{-1}\left(\frac{\boldsymbol{b}\cdot\boldsymbol{c}}{bc}\right), \ \beta = \cos^{-1}\left(\frac{\boldsymbol{a}\cdot\boldsymbol{c}}{ac}\right), \ \gamma = \cos^{-1}\left(\frac{\boldsymbol{a}\cdot\boldsymbol{b}}{ab}\right) \tag{6b}$$

The six lattice parameters (*a, b, c, α, β, γ*) provide a basis-independent description of the unit cell. Conversely, the corresponding real-space lattice vectors can be reconstructed from these parameters. Using a Cartesian reference frame,

$$\mathbf{a}_1 = \begin{pmatrix} a \\ 0 \\ 0 \end{pmatrix}, \tag{7}$$

$$\mathbf{a}_2 = \begin{pmatrix} b\cos\gamma \\ b\sin\gamma \\ 0 \end{pmatrix}, \tag{8}$$

and

$$\mathbf{a}_3 = \begin{pmatrix} c\cos\beta \\ c\dfrac{\cos\alpha - \cos\beta\cos\gamma}{\sin\gamma} \\ c\sqrt{1 - \cos^2\beta - \left(\dfrac{\cos\alpha - \cos\beta\cos\gamma}{\sin\gamma}\right)^2} \end{pmatrix} \tag{9}$$

These reconstructed vectors are assembled into the lattice matrix '*A*', and the fractional atomic coordinates from the CIF are mapped onto the same code-independent structural representation.

Crystallographic symmetry is also considered during reconstruction because a CIF may contain only symmetry-independent atomic sites, while the remaining positions of the unit cell are generated through the associated space-group operations. VQkit therefore uses the available symmetry information to reconstruct the complete periodic atomic arrangement before code-specific input generation. When required, VQkit employs *spglib* to determine the crystallographic symmetry, space group, symmetry operations, and standardized cell representation of the reconstructed structure. This provides a consistent crystallographic basis for subsequent reciprocal-space analysis and high-symmetry *k*-path generation. Following crystallographic reconstruction, the resulting structure can be exported to either supported electronic-structure format. For *VASP*, VQkit generates the corresponding *POSCAR* representation, whereas for *QE* the same structure is expressed through *CELL_PARAMETERS*, *ATOMIC_SPECIES*, and *ATOMIC_POSITIONS*. Thus, both inputs are generated from a single reconstructed crystallographic model.

### 3.2.3 Reciprocal-space sampling and plane-wave parameters

The periodicity of a crystalline solid permits the electronic states to be described in reciprocal space. According to *Bloch*'s theorem, a *Kohn-Sham* state can be written as

$$\psi_{\mathrm{n}\boldsymbol{k}}(\boldsymbol{r}) = e^{\mathrm{i}\boldsymbol{k}\cdot\boldsymbol{r}} u_{\mathrm{n}\boldsymbol{k}}(\boldsymbol{r}) \tag{10}$$

where $n$ denotes the band index, $k$ is the crystal wave vector, and $u_{nk}(r)$ has the periodicity of the real-space lattice. The corresponding *Kohn-Sham* eigenvalues $E_n(k)$ describe the variation of the electronic states throughout reciprocal space.

For a real-space lattice defined by $a_1$, $a_2$, and $a_3$, the reciprocal lattice vectors are

$$b_1 = 2\pi \frac{a_2 \times a_3}{a_1 \cdot (a_2 \times a_3)} \tag{11}$$

$$b_2 = 2\pi \frac{a_3 \times a_1}{a_1 \cdot (a_2 \times a_3)} \tag{12}$$

$$b_3 = 2\pi \frac{a_1 \times a_2}{a_1 \cdot (a_2 \times a_3)} \tag{13}$$

A reciprocal-space vector can therefore be expressed as

$$\boldsymbol{k} = k_1 b_1 + k_2 b_2 + k_3 b_3 \tag{14}$$

where $k_1$, $k_2$, and $k_3$ are the coefficients of $k$ with respect to the reciprocal basis. The use of a well-defined reciprocal primitive basis is important because the numerical coordinates and labels of special $k$-points depend on the chosen lattice representation. In the crystallographic scheme of Hinuma *et al.*, recommended band paths are likewise expressed using $k$-vector coefficients referred to crystallographic primitive reciprocal basis vectors.

Reciprocal-space sampling serves two distinct purposes in electronic-structure calculations. A uniform $k$-point mesh samples the Brillouin zone (BZ) for evaluating quantities such as the charge density, total energy, and forces, whereas a band-structure path samples selected symmetry directions to evaluate the dispersion $E_n(k)$. Accordingly, a typical workflow consists of a self-consistent calculation using a converged uniform $k$-mesh, followed by a band calculation using the converged electronic density along an ordered symmetry path. In VQkit, $k$-path generation therefore refers specifically to construction of the symmetry-dependent path used for band-structure calculations rather than to generation of the uniform *SCF* sampling mesh.

For band-structure calculations, VQkit first analyzes the crystal structure using *spglib*, which identifies the crystallographic symmetry and provides a standardized primitive-cell representation (**Figure 2a** and **b**). The standardized structure is then passed to *SeeK-path*, which generates the recommended high-symmetry reciprocal-space path [28]. The resulting special *k*-points are expressed with respect to the reciprocal basis of the standardized primitive cell and are subsequently converted into the input syntax required by *VASP* or *QE* (**Figure 2c** and **d**). VQkit interpolates the selected number of *k*-points along each path segment and preserves the associated high-symmetry labels for subsequent band-structure visualization.

For a primitive cubic lattice, the recommended high-symmetry path depends on the space-group symmetry. The higher-symmetry cP2 case follows

$$\Gamma-\mathrm{X}-\mathrm{M}-\Gamma-\mathrm{R}-\mathrm{X} \;\mid\; \mathrm{R}-\mathrm{M} \tag{15}$$

whereas the lower-symmetry cP1 case includes the additional M-$X_1$ segment. The vertical bar indicates a discontinuity between path segments. This distinction illustrates why symmetry-dependent *k*-path generation is preferable to applying a single fixed sequence to all primitive cubic structures. Additional symmetry-dependent segments may be required for specific space groups, emphasizing the importance of crystallographically consistent *k*-path generation. VQkit generates these paths in formats compatible with *VASP* and *QE*, with intermediate *k*-points determined by the selected sampling density. The resulting *Kohn-Sham* eigenvalues are plotted against the cumulative reciprocal-space path length to obtain the electronic band dispersion.

In a plane-wave formulation, the periodic part of the *Bloch* state is expanded in reciprocal-lattice vectors G,

$$u_{\mathrm{n}\boldsymbol{k}}(\boldsymbol{r}) = \sum_{\mathrm{G}} c_{\mathrm{n},\boldsymbol{k}+\boldsymbol{G}}\, e^{\mathrm{i}\boldsymbol{G}\cdot\boldsymbol{r}} \tag{16}$$

and the basis is truncated using a kinetic-energy criterion,

$$\frac{\hbar^2}{2m_{\mathrm{e}}}\,|\boldsymbol{k}+\boldsymbol{G}|^2 \leq E_{\mathrm{cut}} \tag{17}$$

where $E_{cut}$ is the plane-wave kinetic-energy cutoff and $m_e$ is the electron mass. Increasing $E_{cut}$ increases the number of plane waves and generally improves basis-set completeness, but also increases computational cost. The plane-wave cutoff and *k*-point sampling density must therefore be converged with respect to the target quantity, such as the total energy, forces, or stress, before production calculations are performed. Within VQkit, these quantities are handled as convergence-controlled numerical parameters so that equivalent input-generation and convergence workflows can be applied to both *VASP* and *QE* while retaining the code-specific syntax required by each electronic-structure package.

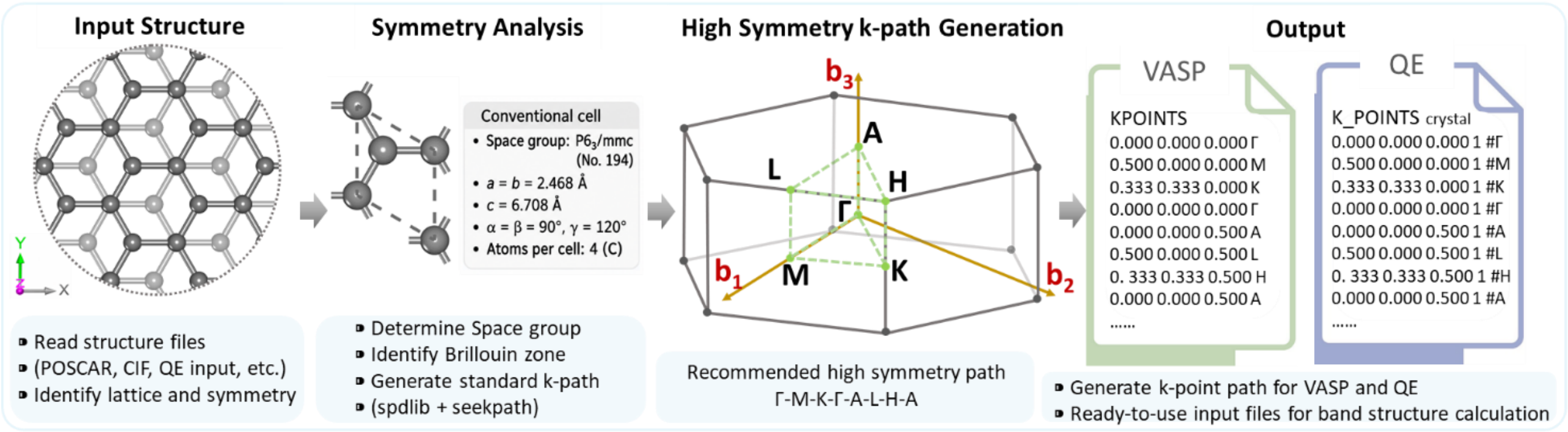

***Figure 2****. VQkit workflow for crystal symmetry analysis and high-symmetry k-path generation, illustrating structure input, space-group identification, Brillouin-zone construction, and automatic generation of KPOINTS files for VASP and QE band-structure calculations.*

## 3.3. Post-processing and Analysis Modules

VQkit provides a unified post-processing framework for extracting and analyzing physical quantities from *VASP* and *QE* calculations. The framework automatically identifies relevant output files and retrieves total energies, Fermi energies, electronic eigenvalues, DOS, forces, convergence information, and optimized structural data. These quantities are processed through common analysis routines, enabling consistent interpretation of results obtained from both electronic-structure codes. Beyond conventional band-structure, DOS, and vibrational-property analyses, VQkit incorporates application-oriented modules for surface properties, reaction energetics, electrocatalysis, atomic dynamics, and mechanical behavior, with representative examples illustrated in **Figure 3a-e**. The mechanical-properties module supports energy-strain and stress-strain approaches for evaluating general elastic tensors, elastic constants, compliance coefficients, Voigt-Reuss-Hill averaged moduli, elastic anisotropy, and mechanical-stability criteria. In the stress-strain approach, VQkit extracts stress tensors from *VASP* and *QE* calculations to reconstruct the corresponding elastic response within the unified analysis framework.

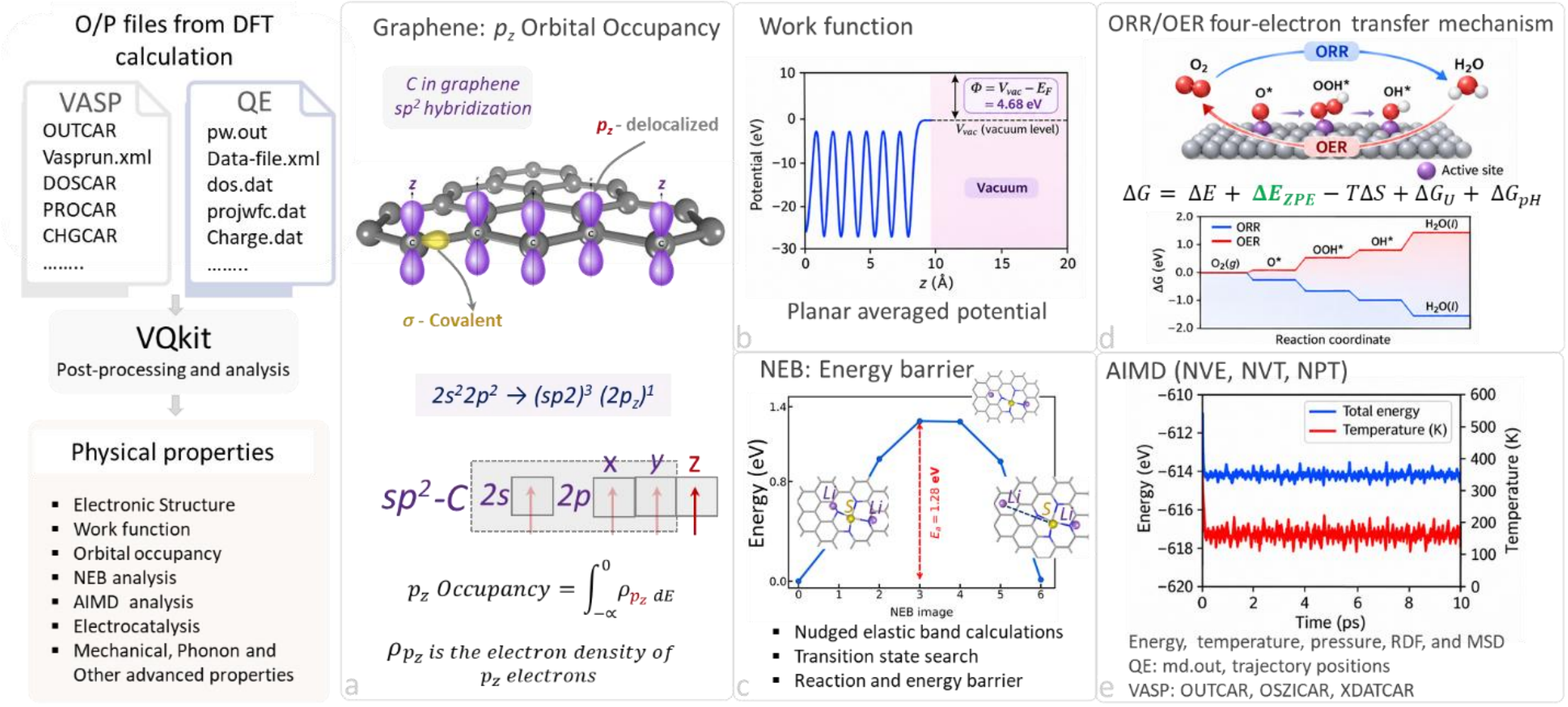

***Figure 3.*** *Overview of VQkit post-processing and materials-property analysis. (a) Graphene ($p_z$)-orbital occupancy and electronic hybridization, (b) work-function analysis from the planar-averaged electrostatic potential, (c) NEB energy-barrier analysis, (d) ORR/OER free-energy pathways, and (e) AIMD analysis. VQkit processes VASP and QE output files to extract electronic, energetic, catalytic, and dynamical properties.*

### 3.3.1. Calculation status and output inspection

Before detailed post-processing, VQkit evaluates whether a calculation has completed successfully and whether the resulting data are suitable for further analysis. For VASP, relevant information is obtained from files such as *OUTCAR*, *OSZICAR*, and *vasprun.xml*, whereas *QE* results are read from the corresponding output file. The module summarizes key quantities, including the final total energy, Fermi energy, number of electronic and ionic iterations, force information, and convergence status. It also detects warning and error indicators associated with incomplete or unconverged calculations. This initial validation reduces manual inspection of lengthy output files and helps prevent subsequent structural or electronic-property analyses from being performed on unreliable results.

### 3.3.2. Electronic band-structure and gap analysis

The electronic band structure contains information beyond the magnitude of the band gap. The dispersion of the *Kohn Sham* eigenvalues, $E_n(\mathbf{k})$, reflects the variation of the electronic states throughout reciprocal space. The gradient of the band energy with respect to the wave vector determines the group velocity,

$$V_n(\mathbf{k}) = \frac{1}{\hbar} \nabla_k E_n(\mathbf{k}), \tag{18}$$

where $V_n(\mathbf{k})$ is the group velocity of an electronic state in band $n$ at the wave vector $\mathbf{k}$, $E_n(\mathbf{k})$ is the corresponding band energy, $\hbar$ is the reduced Planck constant, and $\nabla_k$ denotes the gradient with respect to the reciprocal-space vector **k**. Thus, the slope of the electronic band determines the propagation velocity of the corresponding electronic state.

Near a band extremum, the curvature of the energy dispersion is related to the carrier effective-mass tensor according to

$$\left(m_n^{*\,-1}\right)_{ij} = \frac{1}{\hbar^2}\frac{\partial^2 E_n}{\partial k_i \partial k_j} \tag{19}$$

where $m_n^*$ is the inverse effective-mass tensor associated with band *n*, $i$ and $j$ denote Cartesian reciprocal-space directions, and $k_i$ and $k_j$ are the corresponding components of the wave vector. A strongly curved, highly dispersive band is generally associated with a smaller carrier effective mass, whereas a weakly dispersive or nearly flat band corresponds to a larger effective mass and a greater degree of electronic localization.

Within VQkit, the calculated band energies are referenced to the Fermi level and analyzed along the selected high-symmetry *k*-path. For semiconducting and insulating systems, the valence-band maximum (VBM) and conduction-band minimum (CBM) are identified, and the fundamental band gap is evaluated as

$$E_g = E_{CBM} - E_{VBM} \tag{20}$$

The reciprocal-space positions of the *VBM* and *CBM* are retained to distinguish direct and indirect band gaps. A direct gap occurs when both extrema are located at the same **k** point, whereas an indirect gap occurs when they appear at different reciprocal-space positions. In metallic systems, the intersection of one or more bands with the Fermi level provides the characteristic signature of metallic behavior.

Spin polarization and crystal symmetry provide additional physical information contained in the band dispersion. In collinear spin-polarized calculations, the spin-dependent *Kohn-Sham* potential produces separate spin-up and spin-down electronic eigenvalues. The energy difference between appropriately corresponding spin-resolved states can characterize exchange-induced band splitting, providing information on the spin-dependent electronic structure. However, meaningful comparison requires consideration of the orbital character and symmetry of the corresponding electronic states.

Crystal symmetry independently constrains the allowed electronic eigenstates. At a given wave vector, symmetry operations belonging to the little group of **k** commute with the electronic Hamiltonian, allowing the Bloch states to be classified according to its symmetry representations. Multidimensional representations can enforce degeneracies at particular high-symmetry points or lines, whereas structural or magnetic symmetry breaking can lift these degeneracies and produce characteristic band splitting. The high-symmetry labels and segment boundaries generated through the *spglib/SeeK*-path preprocessing workflow are retained by VQkit during post-processing, allowing *VASP* and *QE* band energies to be plotted using a consistent crystallographic path representation.

### 3.3.3 Mechanical Properties and Elastic Tensor Analysis

For a three-dimensional crystal, the small-strain mechanical response is described by the second-order elastic stiffness tensor[29] . VQkit starts from a fully relaxed reference structure with lattice matrix $A_0$ and volume $V_0$. Its mechanical module implements two independent routes for *VASP* and *QE*: energy-strain curvature fitting and stress-strain slope fitting. The engineering-strain vector and deformed cell are defined as follows

$$\boldsymbol{\eta} = \left(\varepsilon_{xx}, \varepsilon_{yy}, \varepsilon_{zz}, 2\varepsilon_{yz}, 2\varepsilon_{xz}, 2\varepsilon_{xy}\right)^{\mathrm{T}} \tag{21}$$

$$\mathbf{A}(\delta) = [\mathbf{I} + \boldsymbol{\varepsilon}(\delta)]\mathbf{A}_0, \qquad \boldsymbol{\eta}(\delta) = \delta\mathbf{p} \tag{22}$$

Here $\varepsilon$ is the symmetric 3 × 3 strain tensor, $p$ selects a strain mode, and $\delta$ is its scalar amplitude. The factor of two in the last three components of $\eta$ gives the engineering-shear convention. Fractional atomic positions are initially retained when the cell is deformed; atomic coordinates are then relaxed while each strained cell remains fixed. The default grid contains seven amplitudes from -0.015 to +0.015 at intervals of 0.005. These deformations must remain within the harmonic regime, and the reference structure and all strained calculations require consistent, converged electronic settings.

In the energy-strain route, VQkit fits the energy of each strain mode to a quadratic polynomial. The linear term captures residual first-order stress; at a sufficiently relaxed reference, the quadratic curvature gives the zero-pressure stiffness:

$$E_{\mathbf{p}}(\delta) = a_{\mathbf{p}}\delta^2 + b_{\mathbf{p}}\delta + c_{\mathbf{p}}, \qquad q_{\mathbf{p}} = \frac{a_{\mathbf{p}}}{V_0} = \frac{1}{2}\mathbf{p}^{\mathrm{T}}\mathbf{C}\mathbf{p}. \tag{23}$$

Stiffness C and curvature $q$ in this relation have common energy-density units. The *VASP* supplies final electronic free energies (*TOTEN* or *OSZICAR*); *QE* final total energies are converted from *Ry* to *eV*. VQkit converts fitted curvatures from *eV* $Å^{-3}$ to *GPa* using 1 *eV* $Å^{-3}$ = 160.22 *GPa*. For its general crystal model, six modes apply one engineering-strain component at a time and 15 additional modes apply every distinct pair. Denoting their curvatures by $q_i$ and $q_{ij}$, respectively, the 21 independent entries of the symmetric 6 × 6 tensor follow directly:

$$C_{ii} = 2q_i, \qquad C_{ij} = q_{ij} - q_i - q_j \quad (i \neq j) \tag{24}$$

The alternative stress-strain route uses the same six single-component modes. When component 'j' is applied, VQkit fits each of the six computed stress responses to obtain one column of the stiffness matrix:

$$\sigma_i(\delta) = \sigma_i^{(0)} + C_{ij}\delta + O(\delta^2), \qquad C_{ij} = \frac{\mathrm{d}\sigma_i}{\mathrm{d}\delta} \quad (\boldsymbol{\eta} = \delta \mathbf{e}_j) \tag{25}$$

Stress tensors from the final *OUTCAR* or *pw.x* output are converted from *kbar* to *GPa*, reordered to (xx, yy, zz, yz, xz, xy), and given a common tension-positive sign. Independent fits produce a raw 6 × 6 matrix; VQkit reports the maximum mismatch between $C_{ij}$ and $C_{ji}$, symmetrizes the matrix, and optionally projects it onto a supported cubic, hexagonal, tetragonal, or orthorhombic model. With the default grid, the general energy-strain and stress-strain routes prepare 147 and 42 strained calculations, respectively. The stress route therefore uses fewer calculations, but its slopes require particularly well-converged stress tensors.

***Table 1.*** *Comparison of the calculation results on elastic constants and the derived elastic modulus using VASPKIT, VQKit with VASP or QE for HfC, ZrC,* $ZrO_2$*, Ti, and diamond.*

| $C_{ij}$ (GPa) | HfC | | | ZrC | | | $ZrO_2$ | | | Ti | | | Diamond | | |
|---|---|---|---|---|---|---|---|---|---|---|---|---|---|---|---|
| | VASPKIT | VQKit | | VASPKIT | VQKit | | VASPKIT | VQKit | | VASPKIT | VQKit | | VASPKIT | VQKit | |
| | | VASP | QE | | VASP | QE | | VASP | QE | | VASP | QE | | VASP | QE |
| $C_{11}$ | 531.15 | 531.94 | 513.45 | 457.43 | 455.93 | 457.39 | 307.28 | 307.28 | 306.39 | 176.14 | 176.07 | 178.06 | 1050.80 | 1051.19 | 1053.64 |
| $C_{22}$ | 531.15 | 531.94 | 513.45 | 457.43 | 455.93 | 457.39 | 349.66 | 349.66 | 350.79 | 176.14 | 176.07 | 178.06 | 1050.80 | 1051.19 | 1053.64 |
| $C_{33}$ | 531.15 | 531.94 | 513.45 | 457.43 | 455.93 | 457.39 | 256.19 | 258.69 | 255.96 | 188.58 | 188.78 | 188.62 | 1050.80 | 1051.19 | 1053.64 |
| $C_{12}$ | 104.94 | 104.35 | 100.85 | 105.69 | 106.39 | 107.06 | 151.92 | 151.92 | 156.71 | 80.76 | 80.99 | 83.06 | 126.96 | 126.80 | 126.69 |
| $C_{13}$ | 104.94 | 104.35 | 100.85 | 105.69 | 106.39 | 107.06 | 96.18 | 96.93 | 95.35 | 75.22 | 75.21 | 76.19 | 126.96 | 126.80 | 126.69 |
| $C_{23}$ | 104.94 | 104.35 | 100.85 | 105.69 | 106.39 | 107.06 | 141.89 | 141.84 | 148.12 | 75.22 | 75.21 | 76.19 | 126.96 | 126.80 | 126.69 |
| $C_{44}$ | 171.34 | 170.09 | 171.93 | 151.32 | 149.94 | 150.20 | 72.84 | 73.66 | 74.58 | 44.10 | 44.26 | 43.40 | 560.06 | 560.42 | 558.72 |
| $C_{55}$ | 171.34 | 170.09 | 171.93 | 151.32 | 149.94 | 150.20 | 82.57 | 83.15 | 82.82 | 44.10 | 44.26 | 43.40 | 560.06 | 560.42 | 558.72 |
| $C_{66}$ | 171.34 | 170.09 | 171.93 | 151.32 | 149.94 | 150.20 | 116.59 | 116.86 | 117.80 | 47.69 | 47.54 | 47.50 | 560.06 | 560.42 | 558.72 |
| $C_{15}$ | 0.00 | 0.00 | 0.00 | 0.00 | 0.00 | 0.00 | 38.22 | 38.44 | 39.26 | 0.00 | 0.00 | 0.00 | 0.00 | 0.00 | 0.00 |
| $C_{25}$ | 0.00 | 0.00 | 0.00 | 0.00 | 0.00 | 0.00 | -2.30 | -2.44 | -3.69 | 0.00 | 0.00 | 0.00 | 0.00 | 0.00 | 0.00 |
| $C_{35}$ | 0.00 | 0.00 | 0.00 | 0.00 | 0.00 | 0.00 | -3.44 | -3.96 | -2.57 | 0.00 | 0.00 | 0.00 | 0.00 | 0.00 | 0.00 |
| $C_{46}$ | 0.00 | 0.00 | 0.00 | 0.00 | 0.00 | 0.00 | -8.16 | -7.91 | -8.39 | 0.00 | 0.00 | 0.00 | 0.00 | 0.00 | 0.00 |
| **Mechanical properties** | | | | | | | | | | | | | | | |
| $B_V$ | 247.01 | 246.88 | 238.38 | 222.94 | 222.90 | 223.84 | 188.12 | 188.56 | 190.39 | 111.48 | 111.52 | 112.85 | 434.91 | 434.93 | 435.67 |
| $B_R$ | 247.01 | 246.88 | 238.38 | 222.94 | 222.90 | 223.84 | 178.59 | 179.45 | 179.34 | 111.44 | 111.49 | 112.84 | 434.91 | 434.93 | 435.67 |
| $B_H$ | 247.01 | 246.88 | 238.38 | 222.94 | 222.90 | 223.84 | 183.36 | 184.01 | 184.86 | 111.46 | 111.51 | 112.84 | 434.91 | 434.93 | 435.67 |
| $G_V$ | 188.05 | 187.57 | 185.68 | 161.14 | 159.87 | 160.19 | 89.27 | 89.73 | 89.24 | 47.82 | 47.85 | 47.48 | 520.80 | 521.13 | 520.62 |
| $G_R$ | 185.92 | 185.24 | 184.21 | 160.27 | 158.98 | 159.28 | 83.10 | 83.62 | 82.54 | 47.48 | 47.51 | 47.09 | 516.19 | 516.51 | 516.28 |
| $G_H$ | 186.98 | 186.40 | 184.94 | 160.70 | 159.43 | 159.73 | 86.19 | 86.67 | 85.89 | 47.65 | 47.68 | 47.29 | 518.50 | 518.82 | 518.45 |
| $E_H$ | 447.92 | 446.77 | 440.83 | 388.71 | 386.20 | 387.12 | 223.54 | 224.73 | 223.11 | 125.12 | 125.20 | 124.47 | 1113.13 | 1113.65 | 1113.62 |
| $\nu_H$ | 0.1978 | 0.1984 | 0.1918 | 0.2094 | 0.2112 | 0.2118 | 0.2968 | 0.2964 | 0.2989 | 0.3129 | 0.3129 | 0.3162 | 0.0734 | 0.0732 | 0.0740 |
| $B_H/G_H$ | 1.3210 | 1.3245 | 1.2890 | 1.3873 | 1.3982 | 1.4013 | 2.1274 | 2.1230 | 2.1524 | 2.3390 | 2.3386 | 2.3864 | 0.8388 | 0.8383 | 0.8403 |
| $A_U$ | 0.0573 | 0.0630 | 0.0400 | 0.0272 | 0.0282 | 0.0284 | 0.4248 | 0.4164 | 0.4675 | 0.0360 | 0.0355 | 0.0413 | 0.0447 | 0.0447 | 0.0420 |

The compliance tensor S = $C^{-1}$ yields the *Voigt* and *Reuss* estimates of aggregate bulk and shear moduli[30] . To state the implemented expressions compactly, define $C_n = C_{11} + C_{22} + C_{33}$, $C_o = C_{12} + C_{13} + C_{23}$, and $C_s = C_{44} + C_{55} + C_{66}$; $S_n$, $S_o$, and $S_s$ denote the corresponding sums of compliance components. VQkit evaluates

$$K_{\mathrm{V}} = \frac{C_{\mathrm{n}} + 2C_{\mathrm{o}}}{9}, \quad G_{\mathrm{V}} = \frac{C_{\mathrm{n}} - C_{\mathrm{o}} + 4C_{\mathrm{s}}}{15}, \tag{26}$$

$$K_{\mathrm{R}} = \frac{1}{S_{\mathrm{n}} + 2S_{\mathrm{o}}}, \quad G_{\mathrm{R}} = \frac{15}{4(S_{\mathrm{n}} - S_{\mathrm{o}}) + 3S_{\mathrm{s}}} \tag{27}$$

Hill averaging takes the arithmetic mean of each Voigt and Reuss modulus:

$$K_{\mathrm{H}} = \frac{K_{\mathrm{V}} + K_{\mathrm{R}}}{2}, \qquad G_{\mathrm{H}} = \frac{G_{\mathrm{V}} + G_{\mathrm{R}}}{2}. \tag{28}$$

The corresponding isotropic aggregate Young's modulus and Poisson ratio are

$$E_{\mathrm{H}} = \frac{9K_{\mathrm{H}}G_{\mathrm{H}}}{3K_{\mathrm{H}} + G_{\mathrm{H}}}, \qquad \nu_{\mathrm{H}} = \frac{3K_{\mathrm{H}} - 2G_{\mathrm{H}}}{2(3K_{\mathrm{H}} + G_{\mathrm{H}})} \tag{29}$$

The analysis also reports the Hill bulk-to-shear ratio, elastic-anisotropy indices, and eigenvalues of C. Positive eigenvalues indicate a positive-definite small-strain response; cubic Born inequalities are checked whenever the cubic model is used[31] . A default general model imposes no crystal-family equalities, which is appropriate when dopants, defects, or chemical order lower the atomic symmetry of a metrically regular cell. Optional symmetry-reduced models require independent justification: VQkit's automatic family estimate uses lattice lengths and angles, rather than an atomic space-group determination. These reported moduli describe three-dimensional elasticity in *GPa*.

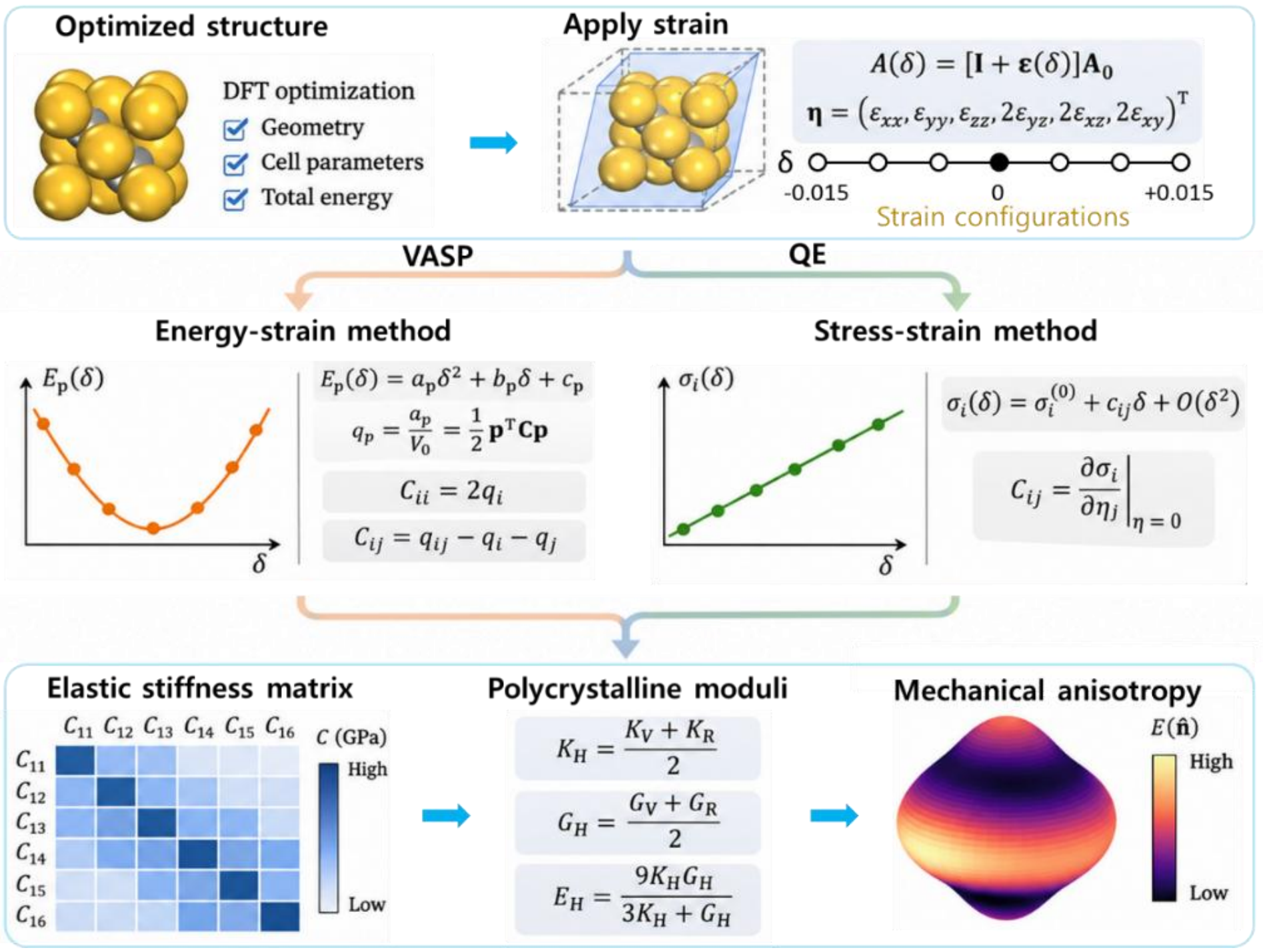

***Figure 4.*** *VQkit workflow for elastic-constant calculations using energy-strain and stress-strain methods, followed by mechanical-property evaluation and three-dimensional elastic anisotropy analysis.*

The post-processing function that constructs the 3-dimensional anisotropy of mechanical properties is also available in VQkit. The loading direction **a** is parameterized by polar angles $\theta$ and $\phi$. A transverse measurement direction **b**, orthogonal to **a**, is parameterized by $\chi$ as in the ELATE treatment[32]:

$$\mathbf{a} = \begin{pmatrix} \sin\theta\cos\phi \\ \sin\theta\sin\phi \\ \cos\theta \end{pmatrix} \tag{30}$$

$$\mathbf{b} = \begin{pmatrix} \cos\theta\cos\phi\cos\chi - \sin\phi\sin\chi \\ \cos\theta\sin\phi\cos\chi + \cos\phi\sin\chi \\ -\sin\theta\cos\chi \end{pmatrix}. \tag{31}$$

The directional responses are evaluated from $S_{ijkl}$ by contraction[33]:

$$E(\mathbf{a}) = \left(S_{ijkl}a_i a_j a_k a_l\right)^{-1}, \qquad \beta(\mathbf{a}) = S_{ijkk}a_i a_j, \tag{32}$$

$$G(\mathbf{a},\mathbf{b}) = \left(4S_{ijkl}a_i b_j a_k b_l\right)^{-1}, \qquad \nu(\mathbf{a},\mathbf{b}) = -\frac{S_{ijkl}b_i b_j a_k a_l}{S_{ijkl}a_i a_j a_k a_l}. \tag{33}$$

Einstein summation applies to Cartesian indices. With $C$ in GPa, $E$ and $G$ are in GPa, $\beta$ is initially linear compressibility in GPa$^{-1}$, and $\nu$ is dimensionless. VQkit multiplies $\beta$ by 1000 for the plotted $TPa^{-1}$ surface, while its directional summary reports $GPa^{-1}$. In the 3D plots, $E$ and $\beta$ each produce one surface. The shear and Poisson surfaces plot $\min_{\chi} X(\mathbf{a}, \mathbf{b})$ and $\max_{\chi} X(\mathbf{a}, \mathbf{b})$ separately for $X = G, \nu$.

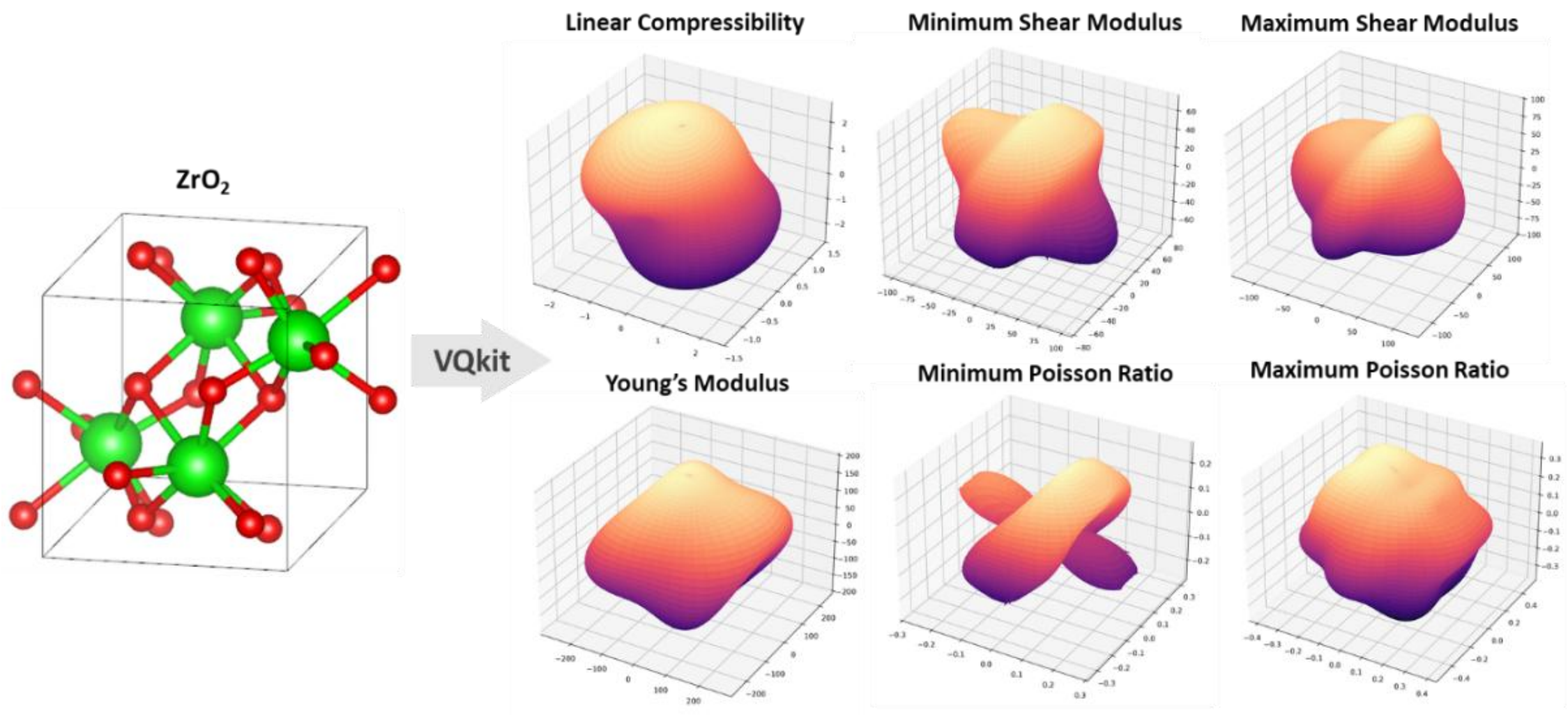

***Figure 5.*** *Three-dimensional representation of mechanical anisotropy generated using VQkit.* $ZrO_2$ *monoclinic crystal structure (left) and corresponding direction-dependent elastic-property surfaces (right) illustrate the variation in mechanical response along different crystallographic directions.*

The example of such anisotropy results on monoclinic $ZrO_2$ is shown in **Figure 5**. The directional properties for linear compressibility, minimum or maximum shear modulus, Young's modulus, and Poisson ratio are calculated from the only required inputs, elastic tensor. The csv files that report the corresponding values are also automatically produced.

### 3.4. Battery-Oriented Analysis of Theoretical Capacity and Voltage

Battery capacity represents the amount of electric charge that can be stored or delivered during electrochemical cycling and is commonly expressed in ampere-hours (*Ah*) or milliampere-hours (*mAh*). The specific capacity is defined as the charge stored or delivered per unit mass of the active electrode material and is expressed in *mAh* $g^{-1}$. Theoretical specific capacity can be estimated from the number of electrons transferred during the electrochemical reaction and the molar mass of the active material.

$$\text{Specific Theoretical capacity } (mAh/g) = \frac{\mathrm{F} \times n_{Li}}{\mathrm{M} \times 3600} \times 1000 \qquad (34)$$

where, F is Faraday's constant (96500 coulombs per mole equivalent), $n_{Li}$ is number of lithium (Li) per formula ($Li_xC_6$) unit of the electrode material. $n_{Li} = x = 6(\frac{n}{n_c})$, $n$ is no of Li atom adsorbed in the anode material, $n_c$ is no of carbon atom in the anode material, M is Molecular mass of the electrode material. Other way,

$$\text{Specific Theoretical capacity } (mAh/g) = \frac{26.801 \times n_{Li} \times 1000}{\text{M}} \tag{35}$$

26.801 = Faraday's constant / (Charging battery/1 h) = 96500/ (3600)

If this has to be expressed in terms of current, divide that by 3600, F = 26.801 Ah Mole$^{-1}$, M = molecular weight for '*mA*' we multiply with 1000.

For graphite anode material, $LiC_6$: $n_{Li} = x = 6(n/n_c) = 6(4/24) = 1$. A complete lithiation corresponds to the formation of $LiC_6$, involving the transfer of one electron per six carbon atoms. Using the molar mass of the graphite host (6 × 12.011= 72.066 g mol$^{-1}$), the theoretical specific capacity is calculated as (1 × 26.801 × 1000)/ 72.066 = 371.9 mAh g$^{-1}$, 371.9 mAh g$^{-1}$, approximately 372 mAh g$^{-1}$ [34]. This value is referenced to the mass of the carbon host, excluding the mass of inserted Li.

The equilibrium voltage can be defined by voltage difference between the two electrodes and also referred to as the open circuit voltage (OCV). As Li is intercalated, there is a change in the electrode's chemical potential that leads to a change in the cell voltage. For full cell, the equilibrium Li intercalation voltage ($\overline{V}$) can be derived by

$$\overline{V} = -\frac{\mu_{Li}^{cathode} - \mu_{Li}^{anode}}{zF} \tag{36}$$

It can calculate by the difference in Li chemical potential ($\mu_{Li}$), between cathode and anode electrode. In the half cell, one of the electrodes is a metallic Li and it is considered as a Li source. Its chemical potential is constant and equal to the Gibbs free energy of Li metal. We calculate average intercalation voltage to lithiate the anode or cathode structure. For a finite amount of reaction, the average intercalation voltage V(x) for various Li content can be expressed a

$$\overline{V}(\text{x}) = -\frac{\Delta G}{zF} \tag{37}$$

where ΔG is the Gibbs free energy of the systems, F is the Faraday constant and *z* is the number of electrons transferred per inserted ion; for $Li^+$, $z = 1$. The anodic cell reaction during Li intercalation can be written as $\text{Li}_{x_1}\text{C}_6 + Li \rightarrow \text{Li}_{x_2}\text{C}_6$. The electrical energy is needed to intercalate Li from $\text{Li}_{x_1}\text{C}_6$ to $\text{Li}_{x_2}\text{C}_6$. Therefore, during the Li intercalation, charge is displaced from Li metal to intercalation compound. Hence, the intercalated energy is equal to the integral of the voltage times the displaced charge [$q = e\,(x_2 - x_1)$].

$$E = \int_{0}^{q_{tot}} V(x)\, \mathrm{dq} \ \text{ or}$$

$$E = -\int_{0}^{q_{tot}} \frac{\mu_{Li}^{IC}(\mathrm{x}) - \mu_{Li}^{0}}{e}\, dq \tag{38}$$

where, $e$ is the electronic charge, $\mu_{Li}^{0}$ and $\mu_{Li}^{IC}$ are the chemical potentials of metallic Li and intercalation compound respectively. Eq. (38) becomes,

$$E = -\int_{x_1}^{x_2} [\mu_{Li}^{IC}(\mathrm{x}) - \mu_{Li}^{0}]\, dx_{Li} \tag{39}$$

$$E = -[G_{Li_{x_2}C_6} - G_{Li_{x_1}C_6} - (x_2 - x_1)G_{Li}] = -\Delta G_r \tag{40}$$

At 0 K, the Gibbs free-energy difference is commonly approximated by the corresponding DFT total-energy difference, neglecting zero-point vibrational contributions. At finite temperatures, pressure-volume, vibrational, configurational, and other entropic contributions may influence the equilibrium voltage. The approximation $\Delta G \approx \Delta E$ is therefore adopted when these contributions are expected to be small relative to the calculated reaction-energy difference.

The average voltage can be determined from the total energy of three structures. There are three independent total energies of $Li_{x_2}C_6$, $Li_{x_1}C_6$ and body centred cubic Li systems are required from DFT calculations. The average voltage is expressed in terms of total energy of the systems involved in the above reaction,

$$\overline{V}(x_1, x_2) = -\frac{[E_{Li_{x_2}C_6} - E_{Li_{x_1}C_6} - (x_2 - x_1)E_{Li}]}{(x_2 - x_1)F} \tag{41}$$

where E($Li_{x_2}C_6$) and E($Li_{x_1}C_6$) are the total energies of the host materials (graphite) with the Li concentrations of $x_2$ and $x_1$. The average Li insertion voltage at $x = ((x_1 + x_2)/2)$ is calculated considering the Li concentration of $x_1$ and $x_2$. This is a valid midpoint used for plotting a voltage interval, but it should not be interpreted as the concentration at which the voltage was independently calculated. However, the real challenge lies in the determination of the relevant thermodynamically stable phases of $Li_{x_1}C_6$ and $Li_{x_2}C_6$.

## 4. Conclusions and Outlook

VQkit provides a unified and extensible computational framework for the pre- and post-processing of density functional theory calculations performed using *VASP* and *QE*. By integrating structure preparation, geometric manipulation, *k*-point generation, input-file construction, calculation inspection, and electronic-structure analysis within a consistent command-line environment, VQkit reduces repetitive manual operations and code-dependent workflow fragmentation. Its shared internal representation enables consistent structural operations and supported post-processing procedures across both electronic-structure packages while retaining their respective input and output conventions. This approach promotes consistent and transparent computational workflows and reduces repetitive code-specific operations in first-principles materials research.

The modular architecture also provides opportunities for further development, including advanced orbital-projected electronic-structure analysis, automated convergence testing, high-throughput processing, enhanced visualization, and computational workflow automation. Overall, VQkit establishes a practical and extensible platform for streamlining first-principles simulation workflows and facilitating consistent computational analysis across *VASP* and *QE*.

## Author contributions

**Rajkamal Anand**: writing original draft, review & editing, validation, visualization conceptualization, formal analysis, data curation, methodology, investigation, supervision. **Seunghwan Kwon**: writing, visualization, validation, formal analysis, and data curation. **Gun Jin Yun**: review & editing, validation, visualization conceptualization, formal analysis, data curation, investigation.

## Declaration of competing interest

Authors declare that they have no known competing financial interests or personal relationships that could have appeared to influence the work reported in this paper.